\documentstyle[12pt]{article}
\begin{document}

\title{On the gravity renormalization off shell.}

\author{~K.~A.~Kazakov,
\thanks{E-mail: $kirill@theor.phys.msu.su$}
P.~I.~Pronin
\thanks{E-mail: $petr@theor.phys.msu.su$}
and K.~V.~Stepanyantz
\thanks{E-mail: $stepan@theor.phys.msu.su$}}

\maketitle

\begin{center}
{\em Moscow State University, Physics Faculty,\\
Department of Theoretical Physics.\\
$117234$, Moscow, Russian Federation}
\end{center}

\begin{abstract}
Using as an example the Einstein gravity with the cosmological
constant, we discuss the calculation of renormalization group functions
off shell. We found, that gauge dependent terms should be absorbed by
the nonlinear renormalization of metric. Nevertheless, some terms can
be included in the renormalization of Newton's constant. This ambiguity
in the renormalization prescription is discussed.
\end{abstract}

\begin{sloppypar}
\noindent
{\bf 1.} The high energy behavior of the gravity interaction draws
the attention of researchers for a very long time. It is well known, that
the perturbation theory for the Einstein gravity differs much from the
Yang-Mills case due to the dimensional coupling constant. Although the
theory is not renormalizable and its contribution to the low energy physics
is very small, a great number of new ideas and field theory methods
originate in the research of the gravity interaction. Here we should
especially mention the background field method \cite{dewitt}. In particular,
it played the important role in the t'Hooft and Veltman derivation of the
algorithm for the one-loop divergences calculation \cite{thooft} (that
allowed to obtain one-loop divergences for Einstein gravity).

The method appeared to be a very powerful calculational tool in
the quantum field theory \cite{leerim,omote}. Its generalizations
\cite{our2} was applied even for the consistency proof of the higher
covariant derivative method \cite{our3}. And nevertheless, there are
some open questions associated with the application of this approach
for the gravity theories. In particular, it is not quite clear why it
is necessary to use motion equations for the renormalization.
The application of the background field method for Yang-Mills theory
or other usual field theory models does not require them at all.
Nevertheless, in the quantum gravity the off shell result is gauge
dependent. For example the special choice of gauge condition in the
Einstein gravity can made the theory finite at the one-loop off shell.
The main reason of motion equation using is that the on shell result
was proven \cite{kallosh2} to be gauge independent. In our opinion
such renormalization is a rather special case. It will be much more
natural to renormalize a theory off shell. Then the natural question is
how to avoid the dependence on the nonphysical parameters?

In this paper on the example of Einstein gravity with cosmological
constant we formulate the prescription for the off shell renormalization
so, that the renormalization of physical values is gauge independent.
Gauge parameters are included in the renormalization of unphysical
metric field and Newton's constant. Using this approach we demonstrate,
that the renormalization of cosmological constant is in a complete agreement
with the on shell results.

Our paper is organized as follows. In next section we calculate the
one-loop counterterms in an arbitrary gauge for the Einstein
gravity with the cosmological constant. The renormalization procedure
off shell is constructed in the section 3. The final section 4 is devoted
to the discussion of the renormalization prescription ambiguity.

\noindent {\bf 2.}
The action for the Einstein gravity with the cosmological constant
has the following form

\begin{equation} S = \frac{1}{{\it k}^2} \int
d^4x\ \sqrt{-g}\ (R-2\Lambda) + \omega \chi
\label{action}
\end{equation}

\noindent
where

\begin{eqnarray}
&&k^2 = 16 \pi G;\nonumber\\
&&R^\sigma _{~\lambda  \mu  \nu } = \partial_\mu \Gamma^\sigma
_{~\lambda \nu }  - \partial_\nu \Gamma^\sigma _{~\lambda \mu } +
\Gamma^\sigma_{~\alpha \mu } \Gamma^\alpha_{~\lambda  \nu } -
\Gamma^\sigma_{~\alpha  \nu }  \Gamma^\alpha_{~\lambda \mu },\\
\end{eqnarray}

\noindent
$\omega$ is the dimensionless coupling constant, $\Lambda$ is
the cosmological constant, $G$ is the Newton's constant and

\begin{equation}
\chi = \frac{1}{32 \pi^2} \int d^4x \sqrt{-g} \left(
R_{\mu \nu \sigma \alpha} R^{\mu \nu \sigma \alpha}
- 4 R_{\mu \nu} R^{\mu \nu} + R^2 \right)
\end{equation}

\noindent
is the Euler number (topological invariant).  The calculation of the
one-loop counterterms can be performed in the framework of the
back\-gro\-und field met\-hod \cite{dewitt}. In accordance with this
method the dynamical field can be rewritten as
$g_{\mu \nu} = g_{\mu \nu} + k h_{\mu \nu}$.

The general coordinates invariance is fixed by adding to the action

\begin{eqnarray}
&&L_{gf}  =  \frac{1}{2} \sqrt{-g} g^{\mu\nu} \chi_\mu \chi_\nu,
\nonumber\\
&&\chi_\mu  =
\frac{1}{\sqrt{1+\alpha}} \left(\nabla_\beta h^\beta_{~\mu} -
\frac{1+\beta}{2} g^{\alpha\beta} \nabla_\mu h_{\alpha\beta} \right)
\label{gauge}
\end{eqnarray}

\noindent
where $\beta$ and $\alpha$ are an arbitrary real constants.

For the quadratic in the quantum fields effective Lagrangian we have

\begin{eqnarray}
L_{eff} & = & - \frac{1}{2} h^{\alpha \beta} \biggl(
{\bf 1}_{\alpha\beta,~\mu\nu} \nabla^2
+ \left(-1+\frac{(1+\beta)^2}{2(1+\alpha)}\right)
g_{\alpha\beta} g_{\mu\nu} \nabla^2
\nonumber \\
&&
+ \frac{2(\alpha-\beta)}{(1+\alpha)}
g_{\alpha\beta}\nabla_\mu \nabla_\nu
- \frac{2 \alpha}{(1+\alpha)} g_{\alpha \mu} \nabla_\beta \nabla_\nu
+ P_{\alpha\beta \mu\nu} \biggr) h^{\mu \nu}
\end{eqnarray}

\noindent
where

\begin{eqnarray}\label{p}
P_{\left((\alpha\beta) (\mu\nu) \right)} & = &
R_{\alpha \mu \beta \nu} - 2 g_{\mu\nu} R_{\alpha \beta}
+ g_{\mu \alpha} R_{\nu \beta}
\nonumber \\
&&
+ \frac{1}{2} g_{\alpha \beta} g_{\mu\nu} \left( R - \Lambda \right)
-\frac{1}{2} \left( R - 2 \Lambda \right) g_{\mu \alpha} g_{\nu \beta}
\end{eqnarray}

\noindent
and

\begin{equation}
{\bf 1}_{\mu\nu,~\alpha\beta} = \frac{1}{2} \left(
g_{\alpha \mu} g_{\beta \nu} +
g_{\alpha \nu} g_{\beta \mu} \right).
\end{equation}

\noindent
The parentheses around couple of indices denote the symmetrization
whereas parenthesis around  four indices means the symmetrization with
pairs' interchange at the same time.

The ghost action obtained in the standard way is
\begin{equation}
L_{gh} = \bar c^\alpha \left( g_{\alpha \beta} \nabla^2
- \beta \nabla_\alpha \nabla_\beta + R_{\alpha \beta}
\right) c^\beta.
\end{equation}

To calculate the one-loop counterterms we use the general expressions
given in \cite{our1,our2} and tensor package \cite{tensor} for the
analytical calculations system REDUCE.  The off-shell one-loop
counterterms including the contributions of both quantum and ghost
fields are

\begin{eqnarray}\label{lambeta}
&&\Gamma^{(1)}_\infty =
\frac{1}{16\pi^2 (d-4)} \int d^4 x
\left(- \frac{58}{5} \Lambda^2
+ \frac{53}{45} \left(
R_{\mu \nu \alpha \beta} R^{\mu \nu \alpha \beta}
- 4 R_{\mu \nu } R^{\mu \nu } + R^2 \right)
\right.
\nonumber\\
&&\left.
\vphantom{\frac{1}{2}}
+ (R-4\Lambda)(a_1 R + a_2 \Lambda)
+ a_3 (10 R_{\mu\nu} R^{\mu\nu}
+ 5 R^2 - 60 \Lambda R + 120 \Lambda^2)
\right)
\end{eqnarray}

\noindent
where

\begin{eqnarray}
&&a_1 = \frac{1}{15}\left(
\alpha (-5\gamma^2 - 10\gamma - 5)
+ 10 \gamma^2 + 5 \gamma - 5
\right),
\nonumber\\
&&a_2 = \alpha (- 2\gamma^3 - 4\gamma^2 - 2\gamma-2)
+ 4\gamma^3 - 6 \gamma - \frac{29}{5},
\nonumber\\
&&a_3 = \frac{1}{60}
\Big(
\alpha^2(\gamma^4 + 4\gamma^3 + 6\gamma^2 + 4\gamma + 4)
+ \alpha (- 4\gamma^4 - 8\gamma^3
\nonumber\\
&&
\vphantom{\frac{1}{2}}
- 2\gamma^2 + 4\gamma + 4)
+ (4\gamma^4 - 6\gamma^2 - 8 \gamma + \frac{21}{5})
\Big).
\end{eqnarray}\label{coef}

\noindent
and we introduced the notation
${\displaystyle \gamma \equiv \frac{\beta}{1-\beta}}$.

In particular, in the case $\Lambda = 0$ this expression is in agreement
with results \cite{kallosh}. The one-loop on-shell counterterms
($R_{\mu\nu} = \Lambda g_{\mu\nu}$) also coincide with the well-known
result \cite{Duff}

\begin{equation} \Gamma^{(1)}_\infty =
\frac{1}{16\pi^2 (d-4)}\ \int d^4 x\
\left( \frac{53}{45} R_{\mu \nu \alpha \beta} R^{\mu \nu \alpha \beta}
- \frac{58}{5} \Lambda^2 \right).
\end{equation}

\noindent
{\bf 3.}
The above calculations show, that the effective action depends on
the gauge parameters. Nevertheless, physical values must be gauge
independent. So, ambiguous terms should be absorbed by the renormalization
of unmeasurable values, for example metric field. For this purpose we will
use the following nonlinear renormalization \cite{thooft,kallosh}

\begin{eqnarray}\label{ren}
&&g_{\mu \nu} \to g_{\mu \nu}^B =  g_{\mu\nu}
+\frac{1}{16\pi^2(d-4)} k^2
\Big(c_1 R g_{\mu \nu} + c_2 \Lambda g_{\mu\nu} + c_3 R_{\mu\nu}\Big);
\nonumber \\
&&\Lambda \to \Lambda_B = \Lambda + c_4\frac{1}{16\pi^2(d-4)} k^2\Lambda^2;
\nonumber \\
&&k^2 \to k^2_B = k^2 + c_5 \frac{1}{16\pi^2(d-4)} k^4 \Lambda.
\end{eqnarray}

Then the bare Lagrangian takes the form

\begin{eqnarray}
&&L(g_{\mu\nu}^B) = L(g_{\mu\nu})
+\frac{1}{16\pi^2(d-4)}\sqrt{-g}
\Big[\left(\frac{1}{2} g^{\mu\nu} (R-2\Lambda) - R^{\mu\nu}\right)
\nonumber\\
&&\times \left(c_1 R g_{\mu\nu} + c_2 \Lambda g_{\mu\nu} + c_3 R_{\mu\nu}
\right) - 2 c_4\Lambda^2 - c_5 \Lambda (R-2\Lambda)\Big]
+ O(R^3).
\end{eqnarray}

$L(g_{\mu\nu}^B)+\Delta L$ should be finite. It leads to the following
equations for the coefficients $c_1 \ldots c_5$:

\begin{eqnarray}
&&- c_3 + 10 a_3=0;\nonumber\\
&&2 c_1 + \frac{1}{2} c_3 + a_1 + 5 a_3 =0;\nonumber\\
&&- 4 c_1 + c_2 - c_3 - c_5 - 4 a_1 + a_2 - 60 a_3=0;\nonumber\\
&&- 4 c_2 - 2 c_4 + 2 c_5 - \frac{58}{5} - 4 a_2 + 120 a_3=0.
\end{eqnarray}

\noindent
They can be rewritten as

\begin{eqnarray}\label{coeff}
&&c_1 = - a_1 - 10 a_3;\nonumber\\
&&c_2 = c_5 - a_2 + 30 a_3; \nonumber\\
&&c_3 = 10 a_3;\nonumber\\
&&c_4 = - \frac{29}{5} - c_5.
\end{eqnarray}

\noindent
{\bf 4.}
(\ref{coeff}) means, there is an ambiguity in the renormalization: gauge
dependent terms can be absorbed in the renormalization either of metric
tensor or of Newton's constant. Is it necessary to find a "true" prescription
of renormalization? We believe, that it is not. Really, the ambiguity does
not affect physical values. The metric field is not measurable, because
motion of a classical particle is completely defined by connection.
As for the Newton's constant, in the considered model it is a pure
multiplicative factor in the Lagrangian, or by the other words an
unessential constant \cite{weinberg}.

Moreover, we are able to avoid the ambiguity by introducing

\begin{eqnarray}\label{new}
&&\lambda= k^2 \Lambda;\nonumber\\
&&G_{\mu\nu} = \frac{1}{k^2}g_{\mu\nu},
\end{eqnarray}

\noindent
so that the Lagrangian will be

\begin{equation}
L = \sqrt{-G}\ (R(G) -2\lambda) + \omega \chi.
\end{equation}

\noindent
(Here $\lambda$ will already be an essential constant.)

The renormalization of $G_{\mu\nu}$ and $\lambda$ does not include
an arbitrary constant as above,

\begin{eqnarray}
&&G_{\mu \nu} \to G_{\mu\nu}
+\frac{1}{16\pi^2(d-4)}
\Big((-a_1-10 a_3) R(G) G_{\mu \nu} +(-a_2+30 a_3) \Lambda G_{\mu\nu}
\nonumber\\
&&+ 10 a_3 R_{\mu\nu}\Big);\nonumber \\
&&\lambda \to \lambda -\frac{29}{5}\frac{1}{16\pi^2(d-4)} \lambda^2.
\end{eqnarray}

\noindent
and coincide with the on shell result.

So, we see, that the ambiguity comes from the fact, that in this
particular model Newton's constant is only multiplicative factor
and is not contained in the motion equations. If matter fields
are added to the Lagrangian, the generalization of (\ref{new}) will
made them dimensionless, for example $\phi \rightarrow \Phi = k \phi$.
Therefore, this substitution allows to avoid the specification of the
mass scale and renormalize only physical dimensionless values.

\end{sloppypar}

\end{document}